\documentclass{aa} 
\usepackage{graphicx}
\usepackage{txfonts}
\begin{document}

\title{The minimum mass for star formation, \\
       and the origin of binary brown dwarfs}

\author{A. P. Whitworth \and D. Stamatellos}

\offprints{A. P. Whitworth}

\institute{School of Physics and Astronomy, Cardiff University,
           5 The Parade, Cardiff CF24 3AA, Wales, UK\\
           \email{anthony.whitworth@astro.cf.ac.uk} \\
           \email{dimitrios.stamatellos@astro.cf.ac.uk}}

\date{Received ; accepted }

\abstract
{$Context:$ The minimum mass for star formation is a critical parameter with profound astrophysical, cosmological and anthropic consequences. Aims. Our first aim is to calculate the minimum mass for $Primary\, F\!ragmentation$ in a variety of potential star-formation scenarios, i.e. (a) hierarchical fragmentation of a 3-D medium; (b) one-shot, 2-D fragmentation of a shock-compressed layer; (c) fragmentation of a circumstellar disc. By Primary Fragmentation we mean fragmentation facilitated by efficient radiative cooling. Our second aim is to evaluate the role of H$_{_2}$ dissociation in facilitating $S\!econdary\, F\!ragmentation$ and thereby producing close, low-mass binaries. $Methods:$ We use power-law fits to the constitutive physics, a one-zone model for condensing fragments, and the diffusion approximation for radiative transport in the optically thick limit, in order to formulate simple analytic estimates.$Results:$ ${\bf (i)}$ For contemporary, local star formation, the minimum mass for Primary Fragmentation is in the range $0.001\,{\rm to}\;0.004\,{\rm M}_{_\odot}$, irrespective of the star-formation scenario considered. This result is remarkable since, both the condition for gravitational instability, and the radiation transport regime operating in a minimum-mass fragment, are different in the different scenarios. $\;\;{\bf (ii)}$ Circumstellar discs are only able to radiate fast enough to undergo Primary Fragmentation in their cool outer parts ($R \ga 100\,{\rm AU}$). Therefore brown dwarfs should have difficulty forming by Primary Fragmentation at $R \la 30\,{\rm AU}$, explaining the Brown Dwarf Desert. Conversely, Primary Fragmentation at $R \ga 100\,{\rm AU}$ could be the source of brown dwarfs in wide orbits about Sun-like stars, and could explain why massive discs extending beyond this radius are rarely seen. $\;\;{\bf (iii)}$ H$_{_2}$ dissociation can lead to collapse and Secondary Fragmentation, thereby converting primary fragments into close, low-mass binaries, with semi-major axes $a \sim 5\,{\rm AU}\,(m_{_{\rm SYSTEM}}/0.1\,{\rm M}_{_\odot})$, in good agreement with observation; in this circumstance, the minimum mass for Primary Fragmentation becomes a minimum $system$ mass, rather than a minimum $stellar$ mass. $\;\;{\bf (iv)}$ Any primary fragment can undergo Secondary Fragmentation, producing a close low-mass binary, provided only that the primary fragment is spinning. Secondary Fragmentation is therefore most likely in primary fragments formed in the outer parts of circumstellar discs (since such fragments inevitably spin), and this could explain why a brown dwarf in a wide orbit about a Sun-like star has a greater likelihood of having a brown-dwarf companion than a brown dwarf in the field -- as seems to be observed. Moreover, we show that binary brown dwarfs formed in this way can sometimes be ejected into the field without breaking up.
}

\maketitle

\section{Introduction}

The structure, physical properties and appearance of the Universe are strongly influenced by the fact that most of the seriously dense baryonic condensations in it are objects with masses in the range $0.01\;{\rm to}\;100\,{\rm M}_{_\odot}$, i.e. stars. By seriously dense we mean
\begin{eqnarray}
\rho & \ga & \rho_{_{\rm B}} \;\sim\; \left[ \frac{m_{_{\rm e}}\,e^2}{2\,\hbar^2\,[4\pi\epsilon_{_0}]} \right]^3\,m_{_{\rm p}} \;\simeq\; 1\,{\rm g}\,{\rm cm}^{-3} \,,
\end{eqnarray}
where $\rho_{_{\rm B}}$ is the critical density separating thermally ionised plasma from pressure-ionised plasma. The Universe would be very different if comparable amounts of baryonic mass had condensed out into seriously dense objects with masses in the ranges $<0.01\,{\rm M}_{_\odot}$ and/or $\,> 100\,{\rm M}_{_\odot}$, but -- as far as we can judge -- they have not. It is therefore appropriate to ask why gravitational condensation, under a wide variety of circumstances, selects objects in the mass range $0.01\;{\rm to}\;100\,{\rm M}_{_\odot}$.

\subsection{The maximum mass for star formation}

The upper mass limit is usually attributed to radiation pressure. As a forming star accumulates matter, its luminosity grows faster than its mass, and eventually -- around $10\,{\rm M}_{_\odot}$, in a spherically symmetric model -- the outward force of radiation pressure becomes greater than the inward force of gravity, making further accretion difficult, but not impossible. More massive stars can form, but their formation requires increasingly extreme and contrived circumstances. One possibility is that the inflow has already acquired such a high ram-pressure before it encounters radiation pressure that it cannot be reversed (McKee \& Tan 2003). Alternatively, if collapse starts from sufficiently dense and homogeneous conditions, the accreting material is already within the dust-destruction radius when the central star switches on, and so the accreting material does not experience the outward force of radiation pressure acting on dust (Edgar \& Clarke 2003). Another possibility is that accretion is highly aspherical (for example, by virtue of being channeled through a disc), and radiation pressure is then released along the directions not occupied by the inflow (for example, along the rotation axis of the disc); this is sometimes called {\it The Flashlight Effect} (Yorke \& Sonnhalter 2002). The final possibility is that stars above $\sim 10\,{\rm M}_{_\odot}$ form by merging of lower-mass stars in the extremely dense and short-lived central cores of forming star-clusters (Bonnell, Bate \& Zinnecker 1998).

\subsection{The minimum mass for star formation}

The lower mass limit is usually attributed to the thermodynamics of interstellar matter (Hoyle 1953; Rees 1976; Low \& Lynden-Bell 1976; Silk 1977; Boss 1988). In order for a condensation to fragment, the Jeans mass in the condensation must continue to decrease, and therefore the gas in the condensation must remain approximately isothermal. This in turn requires that the condensation must radiate away the $PdV$ work done by compression, as fast as it is generated. During the early stages of contraction the rate at which $PdV$ work is done on the condensation (i.e. the compressional heating rate ${\cal H}$) is of order the thermal energy divided by the freefall time. The radiative cooling rate (${\cal L}$) depends on the emissivity and the optical depth of the condensation; it may also depend on how well the gas is thermally coupled to the dust, and on the ambient radiation field. The first concern of this paper is with evaluating the conditions under which ${\cal L} \ga {\cal H}$, and hence the minimum mass for star formation by Primary Fragmentation.

\subsection{The formation of binary systems}

A question closely related to the origin of stars is the origin of their binary properties. The subset of brown dwarfs that are in multiple systems appears to be concentrated in two distinct regions of the parameter-space of multiple systems. If the primary is a Sun-like star, the semi-major axis is usually large, $\ga 100\,{\rm AU}$. If the primary is a brown dwarf, the semi-major axis is usually small, $\la 20\,{\rm AU}$. Moreover, as pointed out by Burgasser, Kirkpatrick \& Lowrance (2005), there seems to be a significant population of hybrid triple or higher-order multiple systems in which a close ($\la 20\,{\rm AU}$) brown dwarf pair  is in a wide orbit ($\ga 100\,{\rm AU}$) around a Sun-like star. The second concern of this paper is to explore a possible explanation for these binary statistics in terms of Secondary Fragmentation, i.e. fragmentation facilitated by H$_{_2}$ dissociation.

\subsection{Plan of the paper}

In Section \ref{SEC:CONSTITUTIVE}, we introduce the constitutive physics which we will use. In Section \ref{SEC:3-D}, we analyse hierarchical fragmentation of a three-dimensional medium, and estimate the resulting minimum mass. In Section \ref{SEC:HIER} we re-evaluate the notion of hierarchical fragmentation, and argue that it probably not a good paradigm for star formation. In Section \ref{SEC:LAYER}, we discuss one-shot fragmentation of a shock-compressed layer, formed by two colliding streams. We derive the resulting minimum mass, and argue that this is a much more realistic scenario for contemporary star formation in turbulent molecular clouds. We also explain why one cannot assume a priori that a minimum-mass fragment is marginally optically thick. In Section \ref{SEC:DISC} we discuss fragmentation of a circumstellar disc and argue that this too may be an important source of low-mass stars, but that it can only be effective at large distances ($R \ga 100\,{\rm AU}$) from the central star. This has important implications for the genesis of exoplanets and brown dwarfs, and for the truncation of massive protostellar discs. In Section \ref{SEC:NONLINEAR} we discuss the nonlinear thermodynamic effects associated with H$_{_2}$ dissociation and the resulting {\it Second Collapse} of a protostar. We explain how these effects may result in the formation of close, low-mass binaries, with separations in good agreement with observation. We also show that close, low-mass binaries could be formed in this way in the outer parts of circumstellar discs and then ejected into the field. In Section \ref{SEC:CONC} we summarise our main conclusions.

\subsection{A note on nomenclature}

We shall assume that, when star-forming gas is approximately isothermal (say ${\rm d}\ell n[T]/{\rm d}\ell n[\rho] \la 0.1$), there are always density fluctuations on a wide range of scales. A density enhancement which might develop into a star is in the first instance termed a {\it proto-fragment}. If a proto-fragment is gravitationally unstable, {\it and} its contraction time-scale is shorter than the timescales of competing processes, it becomes a {\it fragment}. If a fragment can radiate sufficiently fast to stay approximately isothermal it becomes a {\it prestellar  condensation}, and this condition defines the minimum mass for Primary Fragmentation. The implication is that a prestellar condensation is significantly denser (say, at least one hundred times denser) than the initial proto-fragment. A prestellar condensation can subsequently undergo Secondary Fragmentation during The Second Collapse (i.e. when H$_{_2}$ dissociates).

\section{Constitutive physics} \label{SEC:CONSTITUTIVE}

\subsection{Equation of state}

We will be mainly concerned with contemporary star formation in the disc of the Milky Way, and therefore with gas whose chemical composition (by mass) is $70\%$ molecular hydrogen, $28\%$ atomic helium, and $2\%$ heavy elements (in the form of molecules -- like CO -- and dust). The mean gas-particle mass is therefore $\bar{m} \simeq 4.0 \times 10^{-24}\,{\rm g}$, the isothermal sound speed is $a = [k_{_{\rm B}}T/\bar{m}]^{1/2} \simeq 0.06\,{\rm km}\,{\rm s}^{-1}\,[T/{\rm K}]^{1/2}$, and the number density of H$_2$ is $n_{_{{\rm H}_2}} \simeq \rho / [4.8 \times 10^{-24}\,{\rm g}]$. At gas-kinetic temperatures $T \la 100\,{\rm K}$, the rotational levels of H$_{_2}$ are not strongly excited, and therefore the adiabatic exponent is $\gamma \simeq 5/3$. At higher temperatures ($300\,{\rm K}\,\la\,T\,\la\,30,000\,{\rm K}$) $\gamma$ and $\bar{m}$ decrease due to the excitation of the rotational and vibrational levels of H$_{_2}$, the dissociation of H$_{_2}$, and finally the ionisation of H$^{^0}$; we will consider one of the consequences of these changes in Section \ref{SEC:NONLINEAR}.

\subsection{Blackbody fluxes}

For algebraic convenience, we will frequently interchange $T$, $\bar{m}$ and $a=[k_{_{\rm B}}T/\bar{m}]^{1/2}$. In particular, it will sometimes be useful to express the flux from a blackbody surface as
\begin{eqnarray} \label{EQN:BBFLUX}
F_{_{\rm BB}} & \equiv & \sigma_{_{\rm SB}}\,T^4 \;=\; \frac{2\,\pi^5\,\bar{m}^4\,a^8}{15\,c^2\,h^3}\,.
\end{eqnarray}

\subsection{Rosseland- and Planck-mean opacities}

We will also assume that the Rosseland- and Planck-mean opacities due to dust are -- to order of magnitude -- the same, and given by
\begin{eqnarray}
\bar{\kappa}_{_{\rm R}}(T) & \simeq & \bar{\kappa}_{_{\rm P}}(T) \;\,\simeq\,\; \bar{\kappa}_{_{\rm M}}(T) \;\;=\;\; \kappa_{_1}\,\left[\frac{T}{\rm K}\right]^{\,\beta} \,,
\end{eqnarray}
with $\kappa_{_1} = 10^{-3}\,{\rm cm}^2\,{\rm g}^{-1}$ and emissivity index $\beta = 2$ for $T \la 100\,{\rm K}$ (i.e. in the far-infrared and submillimetre wavebands).

\section{Hierarchical Primary Fragmentation of a three-dimensional medium} \label{SEC:3-D}

\subsection{The Jeans Mass in a 3-D medium}

In a uniform 3-D medium, an approximately spherical proto-fragment of mass $m_{_3}$ will only condense out if it is sufficiently massive,
\begin{equation} \label{EQN:MJEANS3}
m_{_3} \;>\; m_{_{\rm JEANS\,3}} 
 \;\simeq\; \left[ \frac{\pi^5\,a^6}{36\,G^3\,\rho} \right]^{1/2} \,,
\end{equation}
or equivalently if it is sufficiently small and dense,
\begin{eqnarray} \label{EQN:RJEANS3}
r_{_3} & < & r_{_{\rm JEANS\,3}} \;\simeq\; 
 \frac{3\,G\,m_{_3}}{\pi^2\,a^2} \,, \\  \label{EQN:RHOJEANS3}
\rho_{_3} & > & \rho_{_{\rm JEANS\,3}}  \;\simeq\; \frac{\pi^5\,a^6}
{36\,G^3\,m_{_3}^2} \,.
\end{eqnarray}
Here the subscript `{\small 3}' records that we are considering a proto-fragment trying to condense out of a 3-D medium.

\subsection{The condensation time-scale for Primary Fragmentation of a 3-D medium}

The timescale on which a Jeans-unstable proto-fragment condenses out of a 3-D medium is given approximately by
\begin{equation} \label{EQN:tCOND3}
t_{_{\rm COND\,3}}(m_{_3}) \simeq \left[\frac{3\,\pi}{32\,G\,\rho}\right]^{1/2} \left\{1 - \left[\frac{m_{_{\rm JEANS\,3}}}{m_{_3}}\right]^{2/3}\right\}^{-1/2} \,,
\end{equation}
i.e. proto-fragments with mass $m_{_3} \gg m_{_{\rm JEANS\,3}}$ condense out on a freefall timescale, whereas less massive proto-fragments take longer, and proto-fragments with mass $m_{_3} \leq m_{_{\rm JEANS\,3}}$ take for ever.

\subsection{Hierarchical Primary Fragmentation of a 3-D medium}

The molecular-cloud gas from which stars are forming today in the Milky Way is expected to be approximately isothermal, with $T \sim 10\,{\rm K}$, as long as it can radiate efficiently via molecular lines and/or dust continuum emission. Therefore it has been argued, following Hoyle (1953), that star formation proceeds in molecular clouds by a process of hierarchical Primary Fragmentation. An initially low-density massive cloud -- which is destined to form a proto-cluster of stars -- satisfies condition (\ref{EQN:MJEANS3}) and starts to contract. Once its density has increased by a factor $f^2$, the Jeans mass $m_{_{\rm JEANS\,3}}$ is reduced by a factor $f^{-1}$, and hence parts of the cloud can condense out independently, thereby breaking up the cloud into $\la f$ sub-clouds. Moreover, as long as the gas remains approximately isothermal (strictly speaking, as long as $a$ remains approximately constant), the process can repeat itself recursively, breaking the cloud up into ever smaller `sub-sub...sub-clouds'.

\subsection{The Opacity Limit for hierarchical Primary Fragmentation of a 3-D medium}

The process ends when the smallest sub-sub...sub-clouds are so optically thick, and/or are collapsing so fast, that the $P\,dV$ work being done on them cannot be radiated away fast enough, and they start to heat up. It is normally presumed that this determines the minimum mass for star formation (e.g. Rees 1976; Low and Lynden-Bell 1976; Silk 1977; Boss 1988), and the termination of hierarchical Primary Fragmentation in this way has traditionally been referred to as {\it The Opacity Limit}. Masunaga and Inutsuka (1999) point out that the gas does not have to be optically thick at The Opacity Limit.

\subsection{The compressional heating rate for a spherical fragment}

To estimate the minimum mass for hierarchical Primary Fragmentation of a 3-D medium, $m_{_{\rm MIN\,3}}$, we first formulate the $P\,dV$ heating rate for a spherical fragment, neglecting the background radiation field,
\begin{eqnarray} \label{EQN:HEAT3}
{\cal H}_{_3} &\equiv& -P_{_3}\,\frac{dV_{_3}}{dt} \,=\, -\,\frac{3m_{_3}a^2}{r_{_3}}\,\frac{dr_{_3}}{dt} \,\sim\, \frac{3m_{_3}a^2}{r_{_3}}\!\left[ \frac{Gm_{_3}}{r_{_3}}\right]^{1/2}\!.
\end{eqnarray}
Here we have obtained the final expression by assuming that the collapse is dynamical and putting $dr_{_3}/dt \sim - [Gm_{_3}/r_{_3}]^{1/2}$.

\subsection{The radiative cooling rate for a spherical fragment}

The maximum radiative luminosity of a spherical fragment is 
\begin{eqnarray} \label{EQN:COOL3}
{\cal L}_{_3} & \;\simeq\; & \frac{16\pi r_{_3}^2\sigma_{_{\rm SB}}T^4}{3\,\left[\bar{\tau}_{_{\rm R}}(T) + \bar{\tau}_{_{\rm P}}^{\;-1}(T)\right]} \,,
\end{eqnarray}
where
\begin{eqnarray} \label{EQN:TAU3}
\bar{\tau}_{_{\rm R}}(T) & \simeq & \bar{\tau}_{_{\rm P}}(T) \;\,\simeq\,\; \frac{3\,m_{_3}\,\bar{\kappa}_{_{\rm M}}(T)}{4\,\pi\,r_{_3}^2}
\end{eqnarray}
are the Rosseland- and Planck-mean optical depths. We emphasise that $T$ is the mean {\em internal} temperature of the fragment, and not necessarily the surface temperature. The treatment of radiation transport in Eqn. (\ref{EQN:COOL3}) is based on the asymptotic forms for radiative diffusion in the optically thick limit ($\bar{\tau}_{_{\rm R}}(T) \gg 1$), and for local emissivity in the optically thin limit ($\bar{\tau}_{_{\rm P}}(T) \ll 1$). It is justified in more detail in Appendix A. Eqn. (\ref{EQN:COOL3}) gives a maximum net luminosity because it neglects the background radiation field (see below).

\subsection{The Opacity Limit for hierarchical Primary Fragmentation of a 3-D medium, assuming $\bar{\tau} \sim 1$}

If we follow Rees (1976) and assume that the fragment is marginally optically thick, we can put $\left[\bar{\tau}_{_{\rm R}}(T) + \bar{\tau}_{_{\rm P}}^{\;-1}(T)\right] \simeq 2$, and then -- using Eqn. (\ref{EQN:BBFLUX}) -- the requirement that ${\cal L}_{_3} \ga {\cal H}_{_3}$ reduces to
\begin{eqnarray} \label{EQN:RCOOL3}
r_{_3} \;\ga\; r_{_{\rm COOL\,3}} & \simeq & \left[\frac{3^6\,5^2}{2^8\,\pi^{12}}\,\frac{G\,m_{_3}^3\,c^4\,h^6}{\bar{m}^8\,a^{12}}\right]^{1/7}\,; \\ \label{EQN:RHOCOOL3}
\rho_{_3} \;\la\; \rho_{_{\rm COOL\,3}} & \simeq & \left[\frac{2^{10}\,\pi^{29}}{3^{11}\,5^6}\,\frac{\bar{m}^{24}\,a^{36}}{G^3\,m_{_3}^2\,c^{12}\,h^{18}}\right]^{1/7}\,.
\end{eqnarray}
This will give a conservative minimum mass, because, in assuming that the fragment is marginally optically thick, we are maximising its luminosity -- all other things ($m_{_3}$, $r_{_3}$ and $T$) being equal.

\subsection{The minimum mass for hierarchical Primary Fragmentation of a 3-D medium, assuming $\bar{\tau} \sim 1$}

Conditions (\ref{EQN:RJEANS3}) and (\ref{EQN:RCOOL3}) require $r_{_{\rm JEANS\,3}} > r_{_{\rm COOL\,3}}$ and hence
\begin{eqnarray} \label{EQN:3DMARG}
m_{_3} \;\ga\; m_{_{\rm MIN\,3}} & \simeq & \left[\frac{5^2\,\pi^2}{2^8\,3}\right]^{1/4}\,\frac{m_{_{\rm PLANCK}}^3}{\bar{m}^2}\,\left[\frac{a}{c}\right]^{1/2}.
\end{eqnarray}
Here $m_{_{\rm PLANCK}} = [hc/G]^{1/2} = 5.5 \times 10^{-5}\,{\rm g}$ is the Planck mass, and $\bar{m}$ cannot differ greatly from $m_{_{\rm p}}$ (the proton mass), so $m_{_{\rm MIN3}}$ is essentially the Chandrasekhar mass ($\sim m_{_{\rm PLANCK}}^3 / m_{_{\rm p}}^2$) times a factor $[a/c]^{1/2} \sim 10^{-3}$. We note the relatively weak dependence of $m_{_{\rm MIN3}}$ on $T$ ($\propto T^{1/4}$) and the relatively strong dependence on $\bar{m}$ ($\propto \bar{m}^{-9/4}$).

For contemporary local star formation, we substitute $\bar{m} \simeq 4.0 \times 10^{-24}\,{\rm g}$ and $a \simeq 1.8 \times 10^4\,{\rm cm}\,{\rm s}^{-1}$ (corresponding to molecular gas at $T \simeq 10\,{\rm K}$) and obtain $m_{_{\rm MIN\,3}} \sim 0.004\,{\rm M}_{_\odot}$.

\subsection{Allowing for radiation transport effects in hierarchical Primary Fragmentation of a 3-D medium}

There is no a priori reason why the limiting fragment should be marginally optically thick (Masunaga \& Inutsuka 1999), and therefore we cannot necessarily put $\left[\bar{\tau}_{_{\rm R}}(T) + \bar{\tau}_{_{\rm P}}^{\;-1}(T)\right] \simeq 2$.

If we assume that the fragment is very optically thin, we must put $\left[\bar{\tau}_{_{\rm R}}(T) + \bar{\tau}_{_{\rm P}}^{\;-1}(T)\right] \simeq \bar{\tau}_{_{\rm P}}^{-1}(T)$, and then after some tedious but straightforward algebra we obtain
\begin{equation} \label{EQN:3DTHIN}
m_{_{\rm MIN\,3}} \;\simeq\; \frac{3^{1/2}\,5}{2^3\,\pi^2}\,\frac{c^2\,h^3}{G\,\bar{m}^4\,a^3\,\bar{\kappa}_{_{\rm M}}(T)} \,.
\end{equation}

Conversely, if we assume that the fragment is very optically thick, we must put $\left[\bar{\tau}_{_{\rm R}}(T) + \bar{\tau}_{_{\rm P}}^{\;-1}(T)\right] \simeq \bar{\tau}_{_{\rm R}}(T)$, and then after some more tedious but straightforward algebra we obtain
\begin{equation} \label{EQN:3DTHICK}
m_{_{\rm MIN\,3}} \;\simeq\; \left[\frac{5\,\pi^4}{2^7\,3^{3/2}}\,\frac{c^2\,h^3\,a^5\,\bar{\kappa}_{_{\rm M}}(T)}{G^5\,\bar{m}^4}\right]^{1/3} \,.
\end{equation}

In general one should use both Eqn. (\ref{EQN:3DTHIN}) and Eqn. (\ref{EQN:3DTHICK}), and then adopt whichever gives the larger value of $m_{_{\rm MIN\,3}}$.

In the case of contemporary local star formation with $T \simeq 10\,{\rm K}$, $\bar{m} \simeq 4.0 \times 10^{-24}\,{\rm g}$, $a \simeq 1.8 \times 10^4\,{\rm km}\,{\rm s}^{-1}$, $\kappa_{_1} \simeq 10^{-3}\,{\rm cm}^2\,{\rm g}^{-1}$ and $\beta \simeq 2$, both expressions (Eqns. \ref{EQN:3DTHIN} and \ref{EQN:3DTHICK}) again give $m_{_{\rm MIN\,3}} \sim 0.004\,{\rm M}_{_\odot}$, but this is because -- coincidentally -- fragments of this mass {\it are} marginally optically thick ($\bar{\tau} \simeq 1$) at their inception.

\section{Problems with hierarchical Primary Fragmentation of a 3-D medium} \label{SEC:HIER}

The above analysis of three-dimensional hierarchical Primary Fragmentation overlooks several effects which all act to increase $m_{_{\rm MIN\,3}}$. Indeed, three-dimensional hierarchical Primary Fragmentation probably does not work. There is no conclusive evidence that hierarchical Primary Fragmentation happens in nature, nor does it seem to occur in numerical simulations of star formation. There are three main reasons.

\subsection{Merging}

First, Eqn. (\ref{EQN:tCOND3}) clearly shows that at each level of the hierarchy, a proto-fragment condenses out more slowly than the larger parent fragment of which it is part, by virtue of the fact that the proto-fragment is always less Jeans unstable than the parent fragment. Therefore a proto-fragment is unlikely to survive as a distinct entity. Instead the overall contraction of the parent fragment will usually cause a proto-fragment to merge with neighbouring proto-fragments.

\subsection{Accretion}

Second, if there are no neighbouring proto-fragments with which to merge, then, even if an isolated proto-fragment starts off with a mass $m_{_3} \sim m_{_{\rm JEANS\,3}}$, it will subsequently grow by a large factor due to accretion. Therefore, before its contraction approaches freefall and it separates out from the background, its final mass will have become much greater than $m_{_{\rm JEANS\,3}}$.

Although the following estimate is only indicative, we adopt the formula for Bondi accretion (Bondi, 1952) . This gives a growth rate
\begin{eqnarray} \nonumber
\frac{dm_{_3}}{dt} &=& \frac{{\rm e}^{3/2}\,\pi\,G^2\,\rho\,m_{_3}^2}{a^3} \\ \label{EQN:BONDI}
&\simeq& \frac{{\rm e}^{3/2}\,\pi^4}{2^3\,6^{1/2}}\,\left[\frac{m_{_3}}{m_{_{\rm JEANS\,3}}}\right]\,\left[\frac{m_{_3}}{t_{_{\rm FF}}}\right] \;\sim\; 22\,\left[\frac{m_{_3}}{m_{_{\rm JEANS\,3}}}\right]\,\left[\frac{m_{_3}}{t_{_{\rm FF}}}\right]\,,
\end{eqnarray}
where $t_{_{\rm FF}} = [3\pi/32G\rho]^{1/2}$ is the freefall time in the background medium of the parent fragment. The implication of Eqns. (\ref{EQN:tCOND3}) and (\ref{EQN:BONDI}) is that a proto-fragment with $m_{_3} \sim m_{_{\rm JEANS\,3}}$ takes several freefall times to condense out, and during one freefall time it increases its mass by more than an order of magnitude.

\subsection{Back-warming}

Third, individual proto-fragments will be back-warmed by the ambient radiation field from other cooling proto-fragments, which in principle fill a significant fraction of the celestial sphere, and again this will reduce their net luminosity ${\cal L}_{_3}$, thereby inhibiting fragmentation.

\section{Two-dimensional one-shot Primary Fragmentation of a shock-compressed layer} \label{SEC:LAYER}

\subsection{Colliding flows, and star formation in a crossing time}

In reality, 3-D hierarchical Primary Fragmentation may be an inappropriate paradigm for star formation in molecular clouds. There is growing evidence that star formation in molecular clouds proceeds very rapidly, essentially `in a crossing time' (Elmegreen 2000), i.e. whilst the molecular cloud is being assembled. In this scenario, star formation occurs in molecular clouds where two turbulent flows of sufficient density collide with sufficient speed -- and therefore sufficient ram pressure -- to produce a gravitationally unstable shock-compressed layer which then fragments to produce prestellar condensations. This mode of Primary Fragmentation is {\it two-dimensional} because the motions which assemble a prestellar condensations out of a shock-compressed layer are, initially, largely in the plane of the layer. It is {\it one-shot} in the sense of not being hierarchical.

\subsection{Geometric considerations}

Strictly speaking, our discussion of shock-compression should not be limited to layers, but should also consider other geometries, in particular filaments and isolated globules. However, we believe that the Primary Fragmentation of a shock-compressed layer captures the essential elements characterising turbulent fragmentation, viz. the dynamic interplay between ram pressure and self gravity in producing prestellar condensations from interstellar gas which is able to keep cool radiatively.

It might be argued that there is limited observational evidence for shock compressed layers, but this is probably a selection effect. Such layers will only be easy to discern under the relatively rare circumstance that they are viewed edge-on --- and then they will be hard to distinguish from filaments.

Moreover, when a shock-compressed layer becomes gravitationally unstable, it fragments first into filaments, and then into individual prestellar condensations (e.g. Turner et al. 1995, Whitworth et al. 1995, Bhattal et al. 1998). The separation between adjacent filaments in the layer is then essentially the same as the separation between prestellar condensations in the same filament (and is given by the fragmentation length, $2 r_{_{\rm FRAG\,2}}$, defined in Eqn. \ref{EQN:RFRAG2} below). This is a generic property of the gravitational fragmentation of a flattened structure, and is also seen in cosmological simulations.

Filaments can also be formed directly if more than two turbulent flows collide more-or-less simultaneously, and this is seen in simulations of interstellar turbulence (e.g. Hennebelle \& Audit 2005; V\'azquez-Semadeni et al. 2006). A filament formed in this way may then be dense enough to fragment gravitationally into a chain of prestellar condensations.

More rarely, convergent turbulent flows may be sufficiently well focussed to form a single compressed prestellar condensation which then collapses in isolation (Whitworth et al. 1996).

However, we will consider here only the case of a shock compressed layer, because it is the simplest case to treat analytically, and probably also the most common case.

\subsection{Growth of a shock-compressed layer}

Specifically, we limit our discussion to the generic case of two flows, both with uniform pre-shock density $\rho$, colliding head-on at relative speed $v$. We assume that the effective isothermal sound speed, $a$, in the post-shock gas of the layer satisfies $a \ll v/2$, or in other words, that the Mach Number of the accretion shock bounding the layer is large:
\begin{eqnarray} \label{EQN:MACH}
{\cal M} & \simeq & \frac{v}{2a} \;\gg\; 1 \,.
\end{eqnarray}
Therefore the surface-density $\Sigma$ of the layer grows according to
\begin{eqnarray} \label{EQN:SIGMA}
\Sigma(t) & \simeq & \rho\,v\,t \,;
\end{eqnarray}
its density is $\;\sim \rho {\cal M}^2 \sim \rho v^2 / 4 a^2$; and its half-thickness is
\begin{eqnarray}
z(t) & \simeq & \frac{2\,a^2\,t}{v} \,.
\end{eqnarray}

\subsection{Structure of a shock-compressed layer}

The layer is contained by the ram pressure of the inflowing gas, rather than its self-gravity, as long as $P_{_{\rm RAM}} \simeq \rho v^2 / 4 \gg G \Sigma^2(t)$, which reduces to
\begin{eqnarray} \label{EQN:RAM}
t & \ll & \frac{1}{2[G\rho]^{1/2}} \,.
\end{eqnarray}
Moreover, there is plenty of time for the layer to relax to hydrostatic equilibrium, as long as its sound crossing time, $t_{_{\rm SC}}(t) \simeq z(t) / a$, obeys the condition $t_{_{\rm SC}}(t) \ll t$. This condition reduces to $v \gg 2a$, and is therefore automatically satisfied (see Eqn. \ref{EQN:MACH}). Thus, until the layer fragments, it has a rather flat density profile, i.e. the density throughout the layer is $\;\sim \rho v^2 / 4 a^2$. 

\subsection{How to estimate the scale on which a shock-compressed layer undergoes Primary Fragmentation}

Before proceeding to estimate the minimum mass for this case ($m_{_{\rm MIN\,2}}$), it is appropriate to sketch the steps which our analysis will take. Because the surface-density of the layer increases monotonically with time (Eqn. \ref{EQN:SIGMA}), the Jeans mass $m_{_{\rm JEANS\,2}}(t)$ (i.e. the minimum mass for a proto-fragment which {\it could} condense out of the layer, {\it given sufficient time}) decreases monotonically with time (Eqn. \ref{EQN:MJEANS2}). Furthermore, it turns out that at time $t$ proto-fragments with mass $m_{_{\rm FASTEST\,2}}(t) \simeq 4 m_{_{\rm JEANS\,2}}(t)$ condense out fastest of all (i.e. faster than both {\it smaller} and {\it larger} proto-fragments), and on a time-scale $t_{_{\rm FASTEST\,2}}(t)$ (Eqn. \ref{EQN:tFASTEST2}). Therefore we hypothesise that non-linear fragmentation of the layer takes place at time $t_{_{\rm FRAG\,2}}$ when the condensation time-scale of the fastest condensing proto-fragment is of order the elapsed time, i.e. $t_{_{\rm FASTEST\,2}}(t_{_{\rm FRAG\,2}}) \simeq t_{_{\rm FRAG\,2}}$ (Eqn. \ref{EQN:tFRAG2}), and this determines the mass of a typical fragment, $m_{_{\rm FRAG\,2}} \simeq m_{_{\rm FASTEST\,2}}(t_{_{\rm FRAG\,2}})$ (Eqn. \ref{EQN:MFRAG2}). We note that this hypothesis has been corroborated by numerical simulations using a variety of numerical codes (Whitworth et al. 1995; Turner et al. 1995; Bhattal et al. 1998; Dale, Bonnell \& Whitworth 2006). Finally, $m_{_{\rm MIN\,2}}$ is the smallest value of $m_{_{\rm FRAG\,2}}$ for which the fragment can radiate sufficiently fast to keep cool as it condenses out.

\subsection{The Jeans mass in a shock-compressed layer}

A disc-like proto-fragment of radius $r_{_2}$, in a shock-compressed layer, evolves according to
\begin{eqnarray} \label{EQN:EVOL2}
\ddot{r}_{_2} & \equiv & \frac{{\rm d}^2r_{_2}}{{\rm d}t^2} \;\simeq\; -\, \pi\,G\,\Sigma(t) \,+\, \frac{a^2}{r_{_2}} \,,
\end{eqnarray}
where the first term on the righthand side of Eqn. (\ref{EQN:EVOL2}) represents self-gravity, and the second term represents hydrostatic support. It follows that at time $t$ only proto-fragments with
\begin{eqnarray} \label{EQN:RJEANS2}
r_{_2} &\ga& r_{_{\rm JEANS\,2}}(t) \,\simeq\, \frac{a^2}{\pi G \Sigma(t)} \,\simeq\, \frac{a^2}{\pi G \rho v t} \,, \\ \nonumber
m_{_2} &\ga& m_{_{\rm JEANS\,2}}(t) \,\simeq\, \pi r_{_{\rm JEANS\,2}}^2(t) \Sigma(t) \,\simeq\, \frac{a^4}{\pi G^2 \Sigma(t)} \,\simeq\, \frac{a^4}{\pi G^2 \rho v t} \,, \\ \label{EQN:MJEANS2}
&& 
\end{eqnarray}
can start to condense out.

Here the subscript `2' records the fact that we are considering a proto-fragment trying to condense out of a 2-D layer.

\subsection{The condensation time-scale for a proto-fragment in a shock-compressed layer}

It is important to note that, in a shock-compressed layer, the Jeans mass is a function of time; it decreases monotonically as the surface density of the layer increases. We now need to evaluate the condensation time-scale for a Jeans-unstable proto-fragment, in order to determine the stage at which such a proto-fragment actually has sufficient time to condense out of the growing layer.

At time $t$, the time-scale on which an unstable proto-fragment condenses out is given by
\begin{eqnarray} \nonumber
t_{_{\rm COND\,2}}(r_{_2},t) & \simeq & \left\{ \frac{r_{_2}}{-\,\ddot{r}_{_2}} \right\}^{1/2} \;\simeq\; \left\{ \frac{\pi\,G\,\Sigma(t)}{r_{_2}} \,-\, \frac{a^2}{r_{_2}^2} \right\}^{-1/2} \\
& \simeq & \frac{a}{\pi G \rho v t} \left\{ \left[ \frac{r_{_{\rm JEANS\,2}}(t)}{r_{_2}}\right] - \left[\frac{r_{_{\rm JEANS\,2}}(t)}{r_{_2}}\right]^2 \right\}^{-1/2} \,;
\end{eqnarray}
or, in terms of $m_{_2}$,
\begin{eqnarray} \label{EQN:tCOND2}
t_{_{\rm COND\,2}}(m_{_2},t) & \simeq & \frac{a}{\pi G \rho v t} \left\{ \left[\frac{m_{_{\rm JEANS\,2}}(t)}{m_{_2}}\right]^{1/2} - \left[\frac{m_{_{\rm JEANS\,2}}(t)}{m_{_2}}\right] \right\}^{-1/2} \,.
\end{eqnarray}

\subsection{The fastest condensing proto-fragment in a shock-compressed layer}

It follows that, at time $t$, the fastest condensing proto-fragment has
\begin{eqnarray}
r_{_{\rm FASTEST\,2}}(t) & \simeq & 2\,r_{_{\rm JEANS\,2}}(t) \,\simeq\, \frac{2 a^2}{\pi G \rho v t} \,, \\
m_{_{\rm FASTEST\,2}}(t) & \simeq & \pi\,r_{_{\rm FASTEST\,2}}^2 \Sigma(t) \,\simeq\, 4 m_{_{\rm JEANS\,2}}(t) \,\simeq\, \frac{4 a^4}{\pi G^2 \rho v t} \,;
\end{eqnarray}
and that the fastest condensing proto-fragment condenses out on a time-scale
\begin{eqnarray} \label{EQN:tFASTEST2}
t_{_{\rm FASTEST\,2}}(t) & \simeq & \frac{2 a}{\pi G \rho v t} \,.
\end{eqnarray}
We note that it is a generic property of the Primary Fragmentation of a layer that there is a finite size of proto-fragment which condenses out faster than both smaller proto-fragments and larger proto-fragments (e.g. Larson 1985).

\subsection{The epoch of non-linear Primary Fragmentation for a shock-compressed layer}

>From Eqn. (\ref{EQN:tFASTEST2}) it follows that the accumulating layer will start non-linear fragmentation once $t_{_{\rm FASTEST\,2}}(t) \la t$, i.e. at
\begin{eqnarray} \label{EQN:tFRAG2}
t & \ga & t_{_{\rm FRAG\,2}} \;\simeq\; \left[ \frac{2\,a}{\pi\,G\,\rho\,v} \right]^{1/2}\,.
\end{eqnarray}

\subsection{The scale of non-linear Primary Fragmentation for a shock-compressed layer} \label{SEC:FRAG2}

Once non-linear Primary Fragmentation occurs in a shock-compressed layer, the characteristic scale of the fragments which form is
\begin{eqnarray} \label{EQN:MFRAG2}
m_{_{\rm FRAG\,2}} &\equiv& m_{_{\rm FASTEST\,2}}(t_{_{\rm FRAG\,2}}) \;\simeq\; \left[ \frac{8\,a^7}{\pi\,G^3\,\rho\,v} \right]^{1/2} \,; \\ \nonumber
 & & \\ \label{EQN:RFRAG2}
r_{_{\rm FRAG\,2}} &\equiv& r_{_{\rm FASTEST\,2}}(t_{_{\rm FRAG\,2}}) \;\simeq\; \left[ \frac{2\,a^3}{\pi\,G\,\rho\,v} \right]^{1/2} \,; \\ \nonumber
 & & \\ \label{EQN:ZFRAG2}
z_{_{\rm FRAG\,2}} &\equiv& z_{_{\rm FASTEST\,2}}(t_{_{\rm FRAG\,2}}) \;\simeq\; \left[ \frac{8\,a^5}{\pi\,G\,\rho\,v^3} \right]^{1/2} \,.
\end{eqnarray}

\subsection{Distinctive features of the Primary Fragmentation of a shock-compressed layer}

We assume that non-linear Primary Fragmentation occurs as soon as it is possible, i.e. at $t \sim t_{_{\rm FRAG\,2}}$, and results in fragments of mass $\sim m_{_{\rm FRAG\,2}}$. This presupposes that there is enough substructure in the pre-shock gas on scales $\; \sim r_{_{\rm FRAG\,2}}$ to seed the fastest condensing proto-fragments. With this assumption, we note the following distinctive aspects of the Primary Fragmentation of a shock-compressed layer.

(i) The characteristic mass, $m_{_{\rm FRAG\,2}}$, and initial radius, $r_{_{\rm FRAG\,2}}$, of a fragment both decrease monotonically with the mass flux, $\rho v$, in the colliding flows.

(ii) Fragments are flattened at their inception, with an aspect ratio approximately equal to the Mach number of the accretion shock bounding the layer:
\begin{eqnarray} \label{EQN:ASPECT}
\frac{r_{_{\rm FRAG\,2}}}{z_{_{\rm FRAG\,2}}} & \simeq & \frac{v}{2\,a} \;\simeq\; {\cal M} \;\gg\; 1 \,.
\end{eqnarray}

(iii) $m_{_{\rm FRAG\,2}}$ is not simply the 3-D Jeans mass ($m_{_{\rm JEANS\,3}}$; Eqn. \ref{EQN:MJEANS3}) evaluated at the post-shock density ($\sim\rho v^2 / 4 a^2$). It is larger by a factor $\sim 12{\cal M}^{1/2}/\pi^3$.

(iv) The shock-compressed layer is still contained by the ram pressure of the inflowing gas, rather than by self-gravity, when its non-linear Primary Fragmentation starts, provided that $t_{_{\rm FRAG\,2}} < 1 / 2 [G\rho]^{1/2}$ (see Eqn. \ref{EQN:RAM}), i.e. provided that ${\cal M} > 4/\pi$, which is satisfied automatically (see Eqn. \ref{EQN:MACH}).

(v) The assumption that matter is still flowing into the accretion shock bounding the layer at time $t_{_{\rm FRAG\,2}}$ requires that the linear extent, $L$, of the pre-shock gas normal to the shock exceed $t_{_{\rm FRAG\,2}} v / 2$, i.e.
\begin{eqnarray}
L & \ga & L_{_{\rm MIN}} \;\simeq\; \left[ \frac{a\,v}{2\,\pi\,G\,\rho} \right]^{1/2} \,.
\end{eqnarray}

(vi) If the pre- and post-shock gases both subscribe to a Larson-type scaling relation, so that the effective sound speed, $a_{_{\rm EFF}}$ (including the contribution from the turbulent velocity dispersion), scales with the mean density, $\bar{\rho}$, according to $a_{_{\rm EFF}} \propto \bar{\rho}\,^{-1/2}$, then the pre-shock gas can have the required extent, $L$, and still be gravitationally stable, provided 
\begin{eqnarray}
L & \la & L_{_{\rm MAX}} \;\simeq\; \frac{a_{_{\rm EFF}}}{[G\,\rho]^{1/2}} \;\simeq\; \left[ \frac{v^2}{4\,G\,\rho} \right]^{1/2} \,.
\end{eqnarray}

(vii) Since $L_{_{\rm MAX}}/L_{_{\rm MIN}} \simeq [\pi {\cal M}/2]^{1/2} > 1$, there is a range of $L$-values for which the pre-shock gas is gravitationally stable, but the shock-compressed layer still undergoes Primary Fragmentation whilst it is accumulating.

(viii) As in  hierarchical Primary Fragmentation, a fragment in a shock-compressed layer will only condense out if it is able to remain approximately isothermal by radiating efficiently.

\subsection{The compressional heating rate for a fragment in a shock-compressed layer}

The compressional heating rate for a flattened disc-like fragment in a layer is
\begin{eqnarray} \label{EQN:HEAT2}
{\cal H}_{_2} &\equiv& -\,P_{_2}\,\frac{dV_{_2}}{dt} \;\simeq\; \frac{\rho\,v^2}{4}\,\frac{2\,\pi\,r_{_2}^2\,z_{_2}}{t_{_{\rm COND\,2}}} \;\sim\; \frac{2a^5}{G}\,,
\end{eqnarray}
where we have obtained the final expression by substituting $r_{_2} = r_{_{\rm FRAG\,2}}$, $z_{_2} = z_{_{\rm FRAG\,2}}$, and $t_{_{\rm COND\,2}} = t_{_{\rm FRAG\,2}}$.

\subsection{The radiative cooling rate for a fragment in a shock-compressed layer}

The radiative cooling rate for a flattened disc-like fragment is
\begin{eqnarray} \label{EQN:COOL2}
{\cal L}_{_2} & \simeq & \frac{4\,\pi\,r_{_2}^2\,\sigma_{_{\rm SB}}\,T^4}{\left[\bar{\tau}_{_{\rm R}}(T)+\bar{\tau}_{_{\rm P}}^{\;-1}(T)\right]} \,, 
\end{eqnarray}
where the optical depths are now given by
\begin{eqnarray} \label{EQN:TAU2}
\bar{\tau}_{_{\rm R}}(T) & \simeq & \bar{\tau}_{_{\rm P}}(T) \;\simeq\; \frac{m_{_2}\,\bar{\kappa}_{_{\rm M}}(T)}{\pi\,r_{_2}^2} \,.
\end{eqnarray}
A detailed justification for Eqns. (\ref{EQN:COOL2}) and (\ref{EQN:TAU2}) is given in Appendix A.

\subsection{The Opacity Limit and the minimum mass for Primary Fragmentation of a shock-compressed layer}

The requirement that ${\cal L}_{_2} \ga {\cal H}_{_2}$ then reduces to a limit on the mass flux, $\rho v$, in the colliding flows,
\begin{equation} \label{EQN:MASSFLUX}
\rho\,v \;\la\; \frac{8\,\pi^5\,\bar{m}^4\,a^6\,/\,15\,c^2\,h^3}
{\left[\bar{\tau}_{_{\rm R}}(T)+\bar{\tau}_{_{\rm P}}^{\;-1}(T)\right]} \,.
\end{equation}

If the fragment is marginally optically thick, we can set $\left[\bar{\tau}_{_{\rm R}}(T)+\bar{\tau}_{_{\rm P}}^{\;-1}(T)\right] \simeq 2$, and then
\begin{equation} \label{EQN:2DMARG}
m_{_2} \;\ga\; m_{_{\rm MIN\,2}} \;\simeq\; \frac{[30]^{1/2}}{\pi^3}\,\frac{m_{_{\rm PLANCK}}^3}{\bar{m}^2} \left[\frac{a}{c}\right]^{1/2}\,.
\end{equation}

If the fragment is optically thin, we obtain
\begin{equation} \label{EQN:2DTHIN}
m_{_{\rm MIN\,2}} \;\simeq\; \frac{15}{4\,\pi^5}\,\frac{c^2\,h^3}{G\,\bar{m}^4\,a^3\,\bar{\kappa}_{_{\rm M}}(T)}\,.
\end{equation}

If the fragment is optically thick, we obtain
\begin{equation} \label{EQN:2DTHICK}
m_{_{\rm MIN\,2}} \;\simeq\; \left[\frac{60}{\pi^7}\,\frac{c^2\,h^3\,a^5\,\bar{\kappa}_{_{M}}(T)}{G^5\,\bar{m}^4}\right]^{1/3} \,.
\end{equation}

In general one should evaluate both Eqn. (\ref{EQN:2DTHIN}) and Eqn. (\ref{EQN:2DTHICK}), and then use whichever gives the larger value of $m_{_{\rm MIN\,2}}$.

In the case of contemporary local star formation with $T \simeq 10\,{\rm K}$, $\bar{m} \simeq 4.0 \times 10^{-24}\,{\rm g}$, $a \simeq 1.8 \times 10^4\,{\rm km}\,{\rm s}^{-1}$, $\kappa_{_1} \simeq 10^{-3}\,{\rm cm}^2\,{\rm g}^{-1}$ and $\beta \simeq 2$, (Eqns. \ref{EQN:2DMARG}, \ref{EQN:2DTHIN} and \ref{EQN:2DTHICK}) all give $m_{_{\rm MIN\,2}} \sim 0.001\,{\rm M}_{_\odot}$, but this is because once again the minimum fragment is -- coincidentally -- marginally optically thick.

We stress that this really is a coincidence. The minimum-mass fragment does not have to be marginally optically thick.  At any temperature, there is a critical opacity,
\begin{eqnarray}
\bar{\kappa}_{_{\rm CRIT}}(T)&\simeq&\frac{15}{(2\pi)^2}\,\left[\frac{G^2\,c^4\,h^6}{\bar{m}\,(k_{_{\rm B}}T)^7}\right]^{1/4}\,.
\end{eqnarray}
Under circumstances where the actual opacity falls below the critical value, $\bar{\kappa}_{_{\rm M}}(T) < \bar{\kappa}_{_{\rm CRIT}}(T)$, the minimum-mass fragment is optically {\it thin} (and cannot cool fast enough because it is optically thin). Conversely, when $\bar{\kappa}_{_{\rm M}}(T) > \bar{\kappa}_{_{\rm CRIT}}(T)$, the minimum-mass fragment is optically {\it thick} (and cannot cool fast enough because it is optically thick); parenthetically, in this regime, somewhat larger fragments which can cool fast enough are also optically thick, but still (approximately) isothermal.

For $T\simeq10\,{\rm K}$ and $\bar{m}\simeq4\times 10^{-24}\,{\rm g}$, we have $\bar{\kappa}_{_{\rm CRIT}}(T=10\,{\rm K})\simeq0.11\,{\rm cm}^2\,{\rm g}^{-1}$, which just happens to be close to our estimate of the opacity in local star forming gas, $\bar{\kappa}_{_{\rm M}}(T=10\,{\rm K})\simeq 0.10\,{\rm cm}^2\,{\rm g}^{-1}$. On the assumption that the amount of dust is proportional to the metallicity, $Z$, but the intrinsic mix of grain properties is universal, we have
\begin{eqnarray}
\frac{\bar{\kappa}_{_{\rm M}}(T)}{\bar{\kappa}_{_{\rm CRIT}}(T)}&\sim&\left[\frac{Z}{0.02}\right]\,\left[\frac{T}{10\,{\rm K}}\right]^{15/4}\,\left[\frac{\bar{m}}{4\times 10^{-24}\,{\rm g}}\right]^{1/4}\,,
\end{eqnarray}
Thus, in regions like the outer parts of the Galaxy, where the metallicity and/or the temperature are lower, the optically thin expression should be valid. Conversely, in regions like the inner parts of the Galaxy, where the metallicity and/or the temperature are higher, the optically thick expression should be valid.

We note also that the minimum mass for two-dimensional, one-shot Primary Fragmentation, $m_{_{\rm MIN\,2}}$ (Eqns. \ref{EQN:2DMARG}, \ref{EQN:2DTHIN} \& \ref{EQN:2DTHICK}), depends on physical variables in exactly the same way as the minimum mass for three-dimensional, hierarchical Primary Fragmentation, $m_{_{\rm MIN\,3}}$ (Eqns. \ref{EQN:3DMARG}, \ref{EQN:3DTHIN} \& \ref{EQN:3DTHICK}). This too appears to be somewhat fortuitous, given that the expression for $m_{_{\rm JEANS\,3}}$ (Eqn. \ref{EQN:MJEANS3}) is quite diferent from that for $m_{_{\rm JEANS\,2}}$ (Eqn. \ref{EQN:MJEANS2}) or $m_{_{\rm FRAG\,2}}$ (Eqn. \ref{EQN:MFRAG2}).

\subsection{Advantages of two-dimensional, one-shot Primary Fragmentation of a layer}

In addition to delivering a smaller minimum mass, two-dimensional one-shot Primary Fragmentation bypasses the three problems associated with three-dimensional hierarchical Primary Fragmentation which we discussed in Section \ref{SEC:HIER}.

First, there is little likelihood of neighbouring fragments merging, since fragments with mass $\sim m_{_{\rm FRAG\,2}}$ condense out of the layer faster than larger structures (see Eqn. (\ref{EQN:tCOND2}) and Larson (1985)).

Second, accretion is less problematic. Boyd \& Whitworth (2004) have analysed in greater detail the radiative cooling of a fragmenting layer, using a two-dimensional model, and taking into account {\it both} on-going accretion (as matter continues to flow into the layer) {\it and} the energy dissipated in the accretion shock. They find that the rate at which energy is dissipated in the accretion shock is comparable with the rate at which the gas in the fragment is heated by compression. For $T \simeq 10\,{\rm K}$, the minimum mass is $\sim 0.0027\,{\rm M}_{_\odot}$. This fragment starts with mass $\sim 0.0011\,{\rm M}_{_\odot}$, but continues to grow by accretion as it condenses out. The minimum mass can be reduced further by decreasing $T$.

Third, there is little backwarming, because there are no other fragments filling the part of the celestial sphere into which a fragment radiates (i.e. perpendicular to the shock-compressed layer).

\section{Primary Fragmentation of a circumstellar disc} \label{SEC:DISC}

Another scenario which may be of more relevance to contemporary star formation than three-dimensional hierarchical Primary Fragmentation is the Primary Fragmentation of a massive circumstellar disc. Massive disc-like structures are observed around some Class 0 and Class I protostars (e.g. Eisner et al. 2005; Eisner \& Carpenter 2006). However, they are quite rare, and we therefore infer, either that discs seldom form, or that they are short-lived. Numerical simulations support the latter inference. For example, in simulations of the collapse and fragmentation of intermediate- and low-mass turbulent cores (Bate et al. 2003; Goodwin et al. 2004), massive disc-like structures form quite commonly around the initial (primary) protostars, but then fragment to form secondary protostars before they can relax to equilibrium. Although massive circumstellar discs may be transient structures, we can only analyse their stability analytically if we assume that they are sufficiently regular to be approximately azimuthally symmetric.

\subsection{The Toomre criterion for Primary Fragmentation of an equilibrium circumstellar disc}

A necessary, but not sufficient, condition for the Primary Fragmentation of an azimuthally symmetric, equilibrium  circumstellar disc is that the surface density $\Sigma$ be sufficiently large,
\begin{equation} \label{EQN:TOOMRE}
\Sigma(R) \;\ga\; \Sigma_{_{\rm TOOMRE}}(R) \;\simeq\; \frac{a(R)\,\epsilon(R)}{\pi\,G}
\end{equation}
(Toomre 1964). Here $R$ is distance from the primary protostar at the centre of the circumstellar disc, $a(R)$ is the local isothermal sound speed,
\begin{eqnarray}
\epsilon(R)&=&\left[R\frac{d\Omega^2}{dR}\,+\,4\Omega^2(R)\right]^{1/2}
\end{eqnarray}
is the local epicyclic frequency\footnote{We have broken with convention in calling the epicyclic frequency $\epsilon$, rather than $\kappa$, simply to avoid confusion with the opacity.}, and $\Omega(R)$ is the orbital angular speed.

\subsection{Criterion for Primary Fragmentation of a non-equilibrium circumstellar disc}

However, since in simulations of star formation circumstellar discs fragment before they have time to approach dynamical equilibrium, it is useful to have a criterion for Primary Fragmentation which can also be applied to non-equilibrium circumstellar discs, where the epicyclic frequency is not properly defined. Such a criterion can be derived, based on the local vorticity, $\omega(R)$, which is defined even for non-equilibrium discs; the only constraint is that the vorticity must be slowly varying on the scale of a proto-fragment, so that it gives a realistic measure of the spin of the proto-fragment. We note that, in the case of a Keplerian disc, the new constraint (Eqn. \ref{EQN:ERMOOT}, below) is exactly equivalent to the old one, because in this circumstance $2\omega(R) = \Omega(R) = \epsilon(R)$. Therefore the results we present subsequently are not affected by which criterion we use (Eqn. \ref{EQN:TOOMRE} or \ref{EQN:ERMOOT}).

Consider a small disc-like proto-fragment of radius $r_{_{\rm D}}$ with its centre at distance $R$ from the centre of the circumstellar disc. Assume that the extent of the disc-like proto-fragment, $\sim 2r_{_{\rm D}}$, is much smaller than the extent of the circumstellar disc of which it is part, $\ga 2R$. Radial excursions of the proto-fragment evolve according to
\begin{eqnarray} \label{EQN:EVOLD}
\ddot{r}_{_{\rm D}} \;\equiv\; \frac{{\rm d}^2r_{_{\rm D}}}{{\rm d}t^2} & \simeq & -\,\pi G \Sigma(R) \,+\, \frac{a^2(R)}{r_{_{\rm D}}} \,+\, \omega^2(R) r_{_{\rm D}} \,.
\end{eqnarray}
In Eqn. (\ref{EQN:EVOLD}), the first two terms on the righthand side represent, respectively, self-gravity and hydrostatic support (just as in Eqn. \ref{EQN:EVOL2}). The third term on the righthand side of Eqn. (\ref{EQN:EVOLD}) represents centrifugal support. The subscript `D' records the fact that we are considering a proto-fragment trying to condense out of a circumstellar disc.

The righthand side of Eqn. (\ref{EQN:EVOLD}) has its minimum when $r_{_{\rm D}} \simeq a(R) / \omega(R)$. This minimum is negative -- signifying instability against contraction -- only if
\begin{eqnarray} \label{EQN:ERMOOT}
\Sigma(R)&\ga&\frac{2\,a(R)\,\omega(R)}{\pi\,G} \,.
\end{eqnarray}
We re-iterate that this criterion for Primary Fragmentation of a general disc (Eqn. \ref{EQN:ERMOOT}) reduces to The Toomre Criterion (Eqn. \ref{EQN:TOOMRE}) if we substitute $2\omega(R) = \epsilon(R)$, as appropriate for a Keplerian disc.

\subsection{The condensation time-scale for a proto-fragment in a circumstellar disc}

>From Eqn. (\ref{EQN:EVOLD}), the condensation time-scale for a proto-fragment in a circumstellar disc is given by
\begin{eqnarray} \nonumber
t_{_{\rm COND\,D}}(R,r_{_{\rm D}}) & \simeq & \left\{ \frac{r_{_{\rm D}}}{-\,\ddot{r}_{_{\rm D}}} \right\}^{1/2} \\ \label{tCONDD}
& \simeq & \left\{ \frac{\pi G \Sigma(R)}{r_{_{\rm D}}} \,-\, \frac{a^2(R)}{r_{_{\rm D}}^2} \,-\, \omega^2(R) \right\}^{-1/2} \,.
\end{eqnarray}

\subsection{The fastest condensing proto-fragment in a circumstellar disc}

$t_{_{\rm COND\,D}}(R,r_{_{\rm D}})$ has a minimum for $r_{_{\rm D}} \simeq 2 a^2(R) / \pi G \Sigma(R)$, and therefore the radius, mass and growth time-scale of the fastest condensing proto-fragment at radius $R$ are given by
\begin{eqnarray} \label{EQN:rFASTESTD}
r_{_{\rm FASTEST\,D}}(R)&\simeq&2\,r_{_{\rm JEANS\,D}}\;\simeq\;\frac{2 a^2(R)}{\pi G \Sigma(R)}\,,\\
m_{_{\rm FASTEST\,D}}(R)&\simeq&4\,m_{_{\rm JEANS\,D}}\;\simeq\;\frac{4\,a^4(R)}{\pi\,G^2\,\Sigma(R)} \, \\ \label{EQN:tFASTESTD}
t_{_{\rm FASTEST\,D}}(R) & \simeq & \left\{ \left[ \frac{\pi G \Sigma(R)}{2 a(R)} \right]^2 \,-\, \omega^2(R) \right\}^{-1/2} \,.
\end{eqnarray}

If we define
\begin{eqnarray}
Q(R) & \equiv & \frac{2\,a(R)\,\omega(R) }{\pi\,G\,\Sigma(R)} \,,
\end{eqnarray}
the condition for Primary Fragmentation of a general disc (Eqn \ref{EQN:ERMOOT}) becomes $Q(R) \la 1$, and Eqns. (\ref{EQN:rFASTESTD}) to (\ref{EQN:tFASTESTD}) reduce to
\begin{eqnarray}
r_{_{\rm FASTEST\,D}}(R) & \simeq & \frac{Q(R)\,a(R)}{\omega(R)} \,, \\
m_{_{\rm FASTEST\,D}}(R) & \simeq & \frac{2\,Q(R)\,a^3(R)}{G\,\omega(R)} \,, \\
t_{_{\rm FASTEST\,D}}(R) & \simeq & \frac{1}{\left[ Q^{-2}(R)\,-\,1 \right]^{1/2}\,\omega(R)} \,.
\end{eqnarray}

\subsection{The typical fragment mass in a dynamically forming circumstellar disc}

Eqn. (\ref{EQN:TOOMRE}) is also approximately the condition for spiral modes to develop in an equilibrium circumstellar disc, and these will have the effect of redistributing angular momentum. As a result, the inner parts of the circumstellar disc may simply accrete onto the central primary protostar, and the outer parts may simply disperse {\it without fragmenting}. The implication is that Primary Fragmentation is much more likely if -- as in the numerical simulations of Bate et al. (2003), Goodwin et al. (2004), Hennebelle et al. (2004) -- the approach to instability is dynamic, rather than quasistatic.

To explore this dynamic situation, we assume that the circumstellar disc grows in mass rather rapidly, and is thereby launched directly into non-linear fragmentation with $Q \sim 0.5$ (rather than $Q$ edging gradually downwards past unity, and instability having first to grow slowly through a linear phase). The radius, mass and condensation time-scale of a typical fragment are then
\begin{eqnarray}
r_{_{\rm FRAG\,D}}(R) & \simeq & \frac{a(R)}{2\,\omega(R)} \,, \\
m_{_{\rm FRAG\,D}}(R) & \simeq & \frac{a^3(R)}{G\,\omega(R)} \,, \\
t_{_{\rm FRAG\,D}}(R) & \simeq & \frac{1}{3^{1/2}\,\omega(R)} \,.
\end{eqnarray}

\subsection{The Gammie criterion for fragmentation of a circumstellar disc}

A second necessary, but not sufficient, condition for the Primary Fragmentation of a circumstellar disc is that fragments be able to cool radiatively on a dynamical time-scale (Gammie 2001); this is just The Opacity Limit under a slightly different guise. Specifically, Gammie suggests that the cooling time-scale for a fragment, $t_{_{\rm COOL}}$, must satisfy $t_{_{\rm COOL}} \la t_{_{\rm ORBIT}}/2$, where $t_{_{\rm ORBIT}}$ is the local orbital period. This condition has been corroborated by numerical simulations (Rice et al. 2003). If a fragment cannot cool sufficiently fast, it is likely to bounce and be sheared apart (e.g. Cai et al., 2006).

\subsection{The compressional heating rate for a fragment in a circumstellar disc}

The compressional heating rate for a fragment at radius $R$ in a circumstellar disc is
\begin{eqnarray}
{\cal H}_{_{\rm D}}(R) & \simeq & \frac{3\,m_{_{\rm FRAG\,D}}(R)\,a^2(R)}{2\,t_{_{\rm FRAG\,D}}(R)} \;\simeq\; \frac{3^{3/2}\,a^5(R)}{2\,G}\,.
\end{eqnarray}

\subsection{The radiative cooling rate for a fragment in a circumstellar disc}

The radiative cooling rate for a fragment in a circumstellar disc is
\begin{eqnarray} \label{EQN:LDISC}
{\cal L}_{_{\rm D}}(R) & \simeq & \frac{4\,\pi\,r_{_{\rm FRAG\,D}}^2(R)\,\sigma_{_{\rm SB}}\,T(R)^4}{\left[\bar{\tau}_{_{\rm R}}(T)+\bar{\tau}_{_{\rm P}}^{\;-1}(T)\right]} \,,
\end{eqnarray}
and the optical depths are given by
\begin{equation} \label{EQN:tauDISC}
\bar{\tau}_{_{\rm R}}(T) \;\simeq\; \bar{\tau}_{_{\rm P}}(T) \;\simeq\; \frac{m_{_{\rm FRAG\,D}}(R)\,\bar{\kappa}_{_{\rm M}}(T)}{\pi\,r_{_{\rm FRAG\,D}}^2(R)} \,.
\end{equation}
These are essentially the same expressions as were invoked for Primary Fragmentation of a shock-compressed layer (Eqns. \ref{EQN:COOL2} and \ref{EQN:TAU2}), since here too we are dealing with a disc-like fragment which radiates mainly through its flat surfaces. See Appendix A for a detailed justification of Eqns. (\ref{EQN:LDISC}) and (\ref{EQN:tauDISC}).

\subsection{The Opacity Limit for Primary Fragmentation of a circumstellar disc}

The requirement that ${\cal L}_{_{\rm D}}(R) \ga {\cal H}_{_{\rm D}}(R)$ therefore reduces to
\begin{equation} \label{EQN:GAMMIE}
\frac{\omega^2(R)}{a^5(R)} \la \frac{2^2\,\pi^6\,G\,\bar{m}^4}{3^{5/2}\,5\,c^2\,h^3\,\left[\bar{\tau}_{_{\rm R}}(T)+\bar{\tau}_{_{\rm P}}^{\;-1}(T)\right]}.
\end{equation}

\subsection{A specific circumstellar-disc model}

To make the discussion more specific we consider a quasi-Keplerian disc around a Sun-like star having mass $M_\star$, luminosity $L_\star$, and hence
\begin{eqnarray}
\omega(R) & \sim & 10^{-7}\,{\rm s}^{-1}\left[\frac{M_{\star}}{{\rm M}_{_\odot}}\right]^{1/2}\left[\frac{R}{\rm AU}\right]^{-3/2} \,, \\ \label{EQN:TDISC}
T(R) & \sim & 300\,{\rm K}\left[\frac{L_{\star}}{{\rm L}_{_\odot}}\right]^{1/4}\left[\frac{R}{\rm AU}\right]^{-1/2} \,, \\
a(R) & \sim & 1\,{\rm km}\,{\rm s}^{-1}\left[\frac{L_{\star}}{{\rm L}_{_\odot}}\right]^{1/8}\left[\frac{R}{\rm AU}\right]^{-1/4} \,.
\end{eqnarray}

The typical fragment at radius $R$ then has radius, mass and optical depth
\begin{eqnarray} \label{EQN:rFRAGD}
r_{_{\rm FRAG\,D}}(R) & \sim & 0.03\,{\rm AU} 
\left[\frac{M_{\star}}{{\rm M}_{_\odot}}\right]^{-1/2} 
\left[\frac{L_{\star}}{{\rm L}_{_\odot}}\right]^{1/8} 
\left[\frac{R}{\rm AU}\right]^{5/4} \,, \\ \label{EQN:mFRAGD}
m_{_{\rm FRAG\,D}}(R) & \sim & 8 \times 10^{-5}\,{\rm M}_{_\odot} 
\left[\frac{M_{\star}}{{\rm M}_{_\odot}}\right]^{-1/2} 
\left[\frac{L_{\star}}{{\rm L}_{_\odot}}\right]^{3/8} 
\left[\frac{R}{\rm AU}\right]^{3/4} \,, \\ \label{EQN:tauFRAGD}
\bar{\tau}_{_{\rm FRAG\,D}}(R) & \sim & 2 \times 10^7 
\left[\frac{M_{\star}}{{\rm M}_{_\odot}}\right]^{1/2} 
\left[\frac{L_{\star}}{{\rm L}_{_\odot}}\right]^{5/8} 
\left[\frac{R}{\rm AU}\right]^{-11/4} \,.
\end{eqnarray}

\subsection{The minimum mass for Primary Fragmentation of a circumstellar disc}
 
>From Eqn. (\ref{EQN:tauFRAGD}) we can assume that the fragment is optically thick, with $\left[\bar{\tau}_{_{\rm R}}(T) + \bar{\tau}_{_{\rm P}}^{\;-1}(T)\right] \simeq \bar{\tau}_{_{\rm FRAG\,D}}(R)$. Eqn. (\ref{EQN:GAMMIE}) then reduces to the form
\begin{eqnarray} \label{EQN:RMIND}
R \;\ga\; R_{_{\rm MIN\,D}} & \sim & 150\,{\rm AU}\,\left[\frac{M_{\star}}{{\rm M}_{_\odot}}\right]^{1/3} \,,
\end{eqnarray}
i.e. prestellar condensations can only form in the outer parts of a circumstellar disc, because only in the outer parts of a disc can such condensations radiate fast enough.

Substituting from Eqn. (\ref{EQN:RMIND}) in Eqns. (\ref{EQN:rFRAGD}), (\ref{EQN:mFRAGD}) and (\ref{EQN:TDISC}), there is a minimum initial condensation radius, a minimum initial condensation mass, and a maximum initial condensation temperature for Primary Fragmentation of a circumstellar disc:
\begin{eqnarray} \label{EQN:rMIND}
r_{_{\rm MIN\,D}} & \sim & 16\,{\rm AU} 
\left[\frac{M_{\star}}{{\rm M}_{_\odot}}\right]^{-1/12} 
\left[\frac{L_{\star}}{{\rm L}_{_\odot}}\right]^{1/8} \,, \\ \label{EQN:mMIND}
m_{_{\rm MIN\,D}} & \sim & 0.003\,{\rm M}_{_\odot} 
\left[\frac{M_{\star}}{{\rm M}_{_\odot}}\right]^{-1/4} 
\left[\frac{L_{\star}}{{\rm L}_{_\odot}}\right]^{3/8} \,, \\ \label{EQN:TMAXD}
T_{_{\rm MAX\,D}} & \sim & 25\,{\rm K} 
\left[\frac{M_{\star}}{{\rm M}_{_\odot}}\right]^{-1/6} 
\left[\frac{L_{\star}}{{\rm L}_{_\odot}}\right]^{1/4} \,.
\end{eqnarray}

Similar conclusions (Eqns. \ref{EQN:RMIND}, \ref{EQN:rMIND}, \ref{EQN:mMIND} and \ref{EQN:TMAXD}) were reached by Rafikov (2005) using closely related arguments.

\subsection{The Brown Dwarf Desert}

Although we have considered a disc with specific surface-density, velocity and temperature profiles, the final result is not very sensitive to these assumptions, as evidenced by the relatively small exponents in Conditions (\ref{EQN:RMIND}), (\ref{EQN:rMIND}) , (\ref{EQN:mMIND}) and (\ref{EQN:TMAXD}). We can allow for some variance in the disc parameters by relaxing these conditions slightly. In particular, we adjust Condition (\ref{EQN:RMIND}) to
\begin{eqnarray}
R&\ga&100\,{\rm AU}\,.
\end{eqnarray}
If brown-dwarf companions to Sun-like stars are formed by Primary Fragmentation of discs, then the constraint that this can only occur at large radii, $R \ga 100\,{\rm AU}$, may explain {\it The Brown-Dwarf Desert}.

\subsection{Dissipation of massive extended protostellar discs}

Observational estimates of the specific angular momentum, $\eta$, in star-forming cores (Bodenheimer, 1995; his Fig. 1) give $\eta \ga 10^{21}\,{\rm cm}^2\,{\rm s}^{-1}$. If deposited in orbit around a $1\,{\rm M}_{_\odot}$ star, this material should end up at radius
\begin{eqnarray}
R&\simeq&\frac{\eta^2}{G{\rm M}_{_\odot}}\;\ga\;300\,{\rm AU}\,.
\end{eqnarray}
It is therefore noteworthy that massive circumstellar discs of this extent are quite rare. The simplest explanation is that they are short-lived, being converted rapidly into brown dwarfs and low-mass H-burning stars, on a dynamical timescale ($\sim 10^4\,{\rm years}$).

\subsection{Forming exoplanets by Primary Fragmentation}

Conversely, our analysis shows that Primary Fragmentation is strongly inhibited at small radii, say $R \la 30\,{\rm AU}$, by the inability of a proto-fragment to radiate fast enough. Therefore it seems very unlikely that gas-giant exoplanets form frequently by Primary Fragmentation. This conclusion is in agreement with the analysis of Rafikov (2005), and with the simulations of Cai et al. (2006). In contrast, Boss (e.g. 2004) has suggested -- again on the basis of numerical simulations -- that gas-giant planets {\it are} able to form by Primary Fragmentation at small radii, because they can cool convectively at these radii. We caution that Boss's interpretation of his numerical results is very speculative. A proto-fragment condensing out gravitationally, {\it on a dynamical timescale}, does not have sufficient time to cool its interior by convection, because this would require the convective cells to migrate and disperse supersonically. What Boss may be seeing is a fragment undergoing an `adiabatic bounce' prior to being sheared apart by differential rotation.

\section{H$_{_2}$ dissociation and Secondary Fragmentation} \label{SEC:NONLINEAR}

\subsection{The spin angular momenta of prestellar condensations}

If low-mass prestellar condensations form by Primary Fragmentation in the outer parts of massive circumstellar discs, in the manner analysed in the preceding section, their subsequent contraction is likely to be moderated by the rate at which they can lose spin angular momentum. This is because, in a marginally unstable disc, there is only a small range of unstable fragment masses. These fragments are only just {\it big} enough for their self-gravity to overcome internal pressure, and only just {\it small} enough for their self-gravity to overcome centrifugal acceleration. Therefore such fragments are naturally born in a state where spin angular momentum makes a significant contribution to their subsequent dynamical evolution.

Moreover, as a prestellar condensation contracts, and its moments of inertia decrease, its ability to loose spin angular momentum to the surroundings by gravitational torques decreases, and the condensation becomes progressively more isolated -- dynamically -- from the rest of the disc. Recent MHD simulations of a protoplanet accreting from a protoplanetary disc (Machida, Inutsuka \& Matsumoto, 2006) show that the protoplanet is able to lose angular momentum by launching bipolar jets. However, this loss of angular momentum does not occur on a dynamical timescale. Therefore we can still presume that condensations forming in discs by Primary Fragmentation derive significant support from rotation, and are quite flattened.

A similar situation holds for low-mass prestellar condensations formed in other ways, for example by hierarchical Primary Fragmentation of a 3-D medium (Section \ref{SEC:3-D}), or by one-shot 2-D Primary Fragmentation of a shock-compressed layer (Section \ref{SEC:LAYER}). All that is required for a condensation to be flattened by rotation is that its specific spin angular momentum, $\ell$, satisfy
\begin{eqnarray} \label{EQN:ETA}
\ell&\sim&G^{1/2}\,m_{_{\rm FRAG}}^{2/3}\,\rho^{-1/6}\,.
\end{eqnarray}
We note that the prestellar condensation can have had a higher $\ell$-value when it first formed, but to reach density $\rho$ without fragmenting, $\ell$ must have been reduced to a value on the order given by Eqn. (\ref{EQN:ETA}).

\subsection{{\it The Second Collapse}}

When the temperature in a prestellar condensation reaches $T_{_{\rm DISS}} \sim 2,000\,{\rm K}$ and the density reaches $\rho_{_{\rm DISS}} \sim 10^{-7}\,{\rm g}\,{\rm cm}^{-3}$, first the vibrational degrees of freedom of H$_{_2}$ start to be excited, and then H$_{_2}$ starts to dissociate. As the temperature continues to increase, more than half the self-gravitational potential energy released by contraction has to be invested in rotational and vibrational degrees of freedom, and then in dissociation of H$_{_2}$, rather than being invested in translational degrees of freedom. As a result the pressure falls below the value required for quasistatic contraction and the condensation collapses (Larson 1969). In the context of isolated protostars, this is normally referred to as {\it The Second Collapse}.

The Second Collapse is further promoted by the fact that these temperatures and densities correspond to {\it The Opacity Gap}, i.e. the regime where dust has sublimated and the H$^{^-}$ ion is not yet abundant, so there is only a very low residual opacity due to molecular lines. As a result, radiative energy leaks out of the condensation very rapidly, accelerating its collapse. However, this is a secondary factor when compared with H$_{_2}$ dissociation (see Whitworth et al. in preparation).

We note that the critical value of the specific angular momentum for H$_{_2}$ dissociation is
\begin{eqnarray}
\ell_{_{\rm DISS}}&\sim&10^{19}\,{\rm cm}^2\,{\rm s}^{-1}\,\left[\frac{m_{_{\rm FRAG}}}{0.1\,{\rm M}_{_\odot}}\right]^{2/3}\,,
\end{eqnarray}
(where we have simply substituted $\rho \sim \rho_{_{\rm DISS}}$ in Eqn. \ref{EQN:ETA}). This is a rather modest specific angular momentum, compared with observational estimates for the gas in star-forming clouds (e.g. Bodenheimer 1995), so the expectation must be that most prestellar condensations are flattened by rotation during their Second Collapse. 

\subsection{Forming close, low-mass binaries by Secondary Fragmentation} \label{SEC:BDBIN1}

If, during The Second Collapse, the spin angular momentum of a prestellar condensation is sufficient to cause it to flatten, it is likely to fragment into two or more pieces (Tsuribe \& Inutsuka, 1999a, 1999b), and hence to form a binary system, or an unstable higher-order multiple system. We call this Secondary Fragmentation. The mean semi-major axis of a binary system formed in this way is of order 
\begin{eqnarray} \label{EQN:aSM1}
a&\sim&\frac{G\,\bar{m}\;m_{_{\rm FRAG}}}{3\,k_{_{\rm B}}\,T_{_{\rm DISS}}}\;\sim\;50\,{\rm AU}\,\left[\frac{m_{_{\rm FRAG}}}{{\rm M}_{_\odot}}\right]\,.
\end{eqnarray}
Higher-order multiples are likely to evolve dynamically by ejecting lower-mass components, and this may ultimately lead to somewhat closer binary systems than predicted by Eqn. (\ref{EQN:aSM1}).

In discussing binary systems with brown-dwarf primaries, it is standard practice to include also systems with very low-mass hydrogen burning primaries $m_{_1}\la\,0.1\,{\rm M}_{_\odot}$. These systems will therefore have semi-major axes 
\begin{eqnarray} \label{EQN:aSM2}
a&\la&5\,{\rm AU}\,\left[\frac{(m_{_1}+m_{_2})}{0.1\,{\rm M}_{_\odot}}\right]\,.
\end{eqnarray}
Here we have assumed that most of the mass of the original prestellar condensation (the one formed by Primary Fragmentation) goes into the two components of the binary, with masses $m_{_1}$ and $m_{_2}$. This is a reasonable assumption, given that the prestellar condensation is by this stage very tightly bound. The inequality in Eqn. (\ref{EQN:aSM2}) acknowledges the possibility that the binary is hardened, by dynamical ejection of additional components in an unstable higher-order multiple, and/or by dissipative interactions with residual gas.

Comparing the locus described by Eqn. (\ref{EQN:aSM1}) with the observations reported in Close et al. (2003; their Fig. 15), and noting the good agreement, we speculate that the genesis of close, low-mass binaries is due to The Second Collapse. Confirmation of this suggestion will require detailed numerical simulations, with appropriate initial and boundary conditions, realistic thermodynamics, and physical transport of angular momentum. Bate (1998) reports a simulation of Secondary Collapse, which fails to result in Secondary Fragmentation. However, this simulation uses a barotropic equation of state, rather than a proper treatment of the energy equation and the associated radiation transport. It may also be influenced by numerical transport of angular momentum. We -- and others -- are currently revisiting this problem using an SPH code with radiation transport, but conclusive results are not yet available.

\subsection{Forming close, low-mass binaries by Secondary Fragmentation in circumstellar discs} \label{SEC:BDBIN2}

Secondary Fragmentation requires that the primary fragment be rotating, and the only Primary Fragmentation scenario in which rotation is explicitly considered is disc fragmentation. Therefore we now consider whether close low-mass binaries can be formed in the outer parts of massive extended discs, {\it and} whether close low-mass binaries formed in this location may subsequently end up in the field.

We note that a close, low-mass binary formed in this way has binding energy
\begin{eqnarray}
-\,\Omega_{_{\rm CLOSE}}&=&\frac{G\,m_{_1}\,m_{_2}}{2\,a}\\
&\ga&6\times 10^{42}\,{\rm erg}\,\left[\frac{(m_{_1}+m_{_2})}{0.1\,{\rm M}_{_\odot}}\right]\,,
\end{eqnarray}
where we have assumed equal-mass components ($m_{_1} = m_{_2}$) and used Eqn. (\ref{EQN:aSM2}). For comparison, the binding energy of the close, low-mass binary as a whole to the Sun-like star at the centre of the original disc is 
\begin{eqnarray}
-\,\Omega_{_{\rm WIDE}}&=&\frac{G\,(m_{_1}+m_{_2})\,M_\star}{R}\\
&\sim&20\times10^{42}\,{\rm erg}\,\left[\frac{(m_{_1}+m_{_2})}{0.1\,{\rm M}_{_\odot}}\right]\,\left[\frac{M_\star}{{\rm M}_{_\odot}}\right]\,\left[\frac{R}{100\,{\rm AU}}\right]^{-1}\,.
\end{eqnarray}
It may therefore be possible for tidal encounters occasionally to detach a close, low-mass binary from a central Sun-like star, without destroying it (i.e. to unbind the wide system without unbinding the close one), and thereby to populate the field with close, low-mass binaries that were originally formed in circumstellar discs. 

To evaluate this possibility more accurately, we consider a perturbing star of mass $M_{_{\rm PERT}}$, travelling at speed $v_{_{\rm PERT}}$ and passing a binary system with total mass $m_{_{\rm BIN}}$ and separation $s_{_{\rm BIN}}$, with the distance of closest approach being $D_{_{\rm PERT}}$. The tide of the passing star will deliver a velocity impulse to one component of the binary, relative to the other, given by the product of the tidal acceleration and the duration of the interaction,
\begin{eqnarray} \label{EQN:Deltav}
\Delta v&\sim&\frac{G\,M_{_{\rm PERT}}\,s_{_{\rm BIN}}}{D_{_{\rm PERT}}^3}\,\times\,\frac{D_{_{\rm PERT}}}{v_{_{\rm PERT}}}\,;
\end{eqnarray}
strictly speaking, the duration of the interaction (the second term on the righthand side of Eqn. \ref{EQN:Deltav}) should be
\[{\rm MIN}\left(\frac{D_{_{\rm PERT}}}{v_{_{\rm PERT}}},\,\left[\frac{s_{_{\rm BIN}}^3}{Gm_{_{\rm BIN}}}\right]^{1/2}\right)\,,
\]
but including this factor would only strengthen our conclusion, so we omit it in the interests of simplicity. The velocity impulse given by Eqn. (\ref{EQN:Deltav}) will unbind the binary if it exceeds
\begin{eqnarray}
v_{_{\rm ESC}}&\sim&\left[\frac{G\,m_{_{\rm BIN}}}{s_{_{\rm BIN}}}\right]^{1/2}\,.
\end{eqnarray}
Therefore disruption requires 
\begin{eqnarray}
\frac{m_{_{\rm BIN}}}{s_{_{\rm BIN}}^3}&\la&\frac{G\,M_{_{\rm PERT}}^2}{D_{_{\rm PERT}}^4\,v_{_{\rm PERT}}^2}\,.
\end{eqnarray}
Using the parameters derived above, we estimate $m_{_{\rm BIN}}/s_{_{\rm BIN}}^3 \ga 0.5\,\times 10^{-7}\,{\rm g}\,{\rm cm}^{-3}$ for the typical close, low-mass binary formed as a result of The Second Collapse, whereas $m_{_{\rm BIN}}/s_{_{\rm BIN}}^3 \la 0.5\times 10^{-12}\,{\rm g}\,{\rm cm}^{-3}$ for a typical wide binary formed as a result of disc fragmentation. Therefore there appears to be a signficiant range of perturber parameters which allow the close, low-mass system to survive whilst destroying the wide system.

Thus it is possible that close, low-mass binaries are formed in the outer parts of massive circumstellar discs round Sun-like stars, in a two-stage process: (i) the formation of a low-mass prestellar condensation by Primary Fragmentation (facilitated by the fact that the condensation can radiate sufficiently fast through thermal dust emission to keep cool); (ii) Secondary Fragmentation of this primary fragment (during The Second Collapse, i.e. whilst H$_{_2}$ is dissociating) to produce a close, low-mass binary system. Additionally, it may sometimes be possible to detach a close, low-mass binary formed in this way from the Sun-like star at the centre of the natal disc, without destroying the close, low-mass binary system -- and thereby to populate the field with close, low-mass binaries.

This scenario is attractive because it explains why Burgasser et al. (2005) infer that brown dwarfs in wide orbits around Sun-like stars have a higher probability of being in a close binary system with another brown dwarf than do brown dwarfs in the field. However, we must caution that this inference is presently based on small-number statistics and needs to be confirmed with a larger sample.

\section{Conclusions} \label{SEC:CONC}

We have reviewed the thermodynamic processes which are presumed to determine the minimum mass for star formation -- namely (i) Primary Fragmentation and the Opacity Limit, (ii) Secondary Fragmentation during H$_{_2}$ dissociation -- and the consequences these thermodynamic processes have for the statistics of binary systems containing brown dwarfs.

\subsection{Primary Fragmentation and the Opacity Limit}

Here the presumption is that a prestellar condensation only fragments if it can radiate fast enough to stay approximately isothermal. We have treated three different generic star formation scenarios, (a) hierarchical fragmentation of a 3-D medium, (b) one-shot 2-D fragmentation of a shock-compressed layer, and (c) fragmentation of a circumstellar disc. We believe that these three scenarios cover the different basic situations in which minimum-mass fragments may be created (as regards geometry, background dynamics, competing processes, etc.).

Hierarchical 3-D fragmentation represents fragmentation in a cloud undergoing overall collapse, with no preferred direction (i.e. a statistically isotropic situation). The main processes competing (successfully) against fragmentation are contraction of the background and accretion. We re-derive the expression for the minimum mass obtained by Rees (1976) under the assumption that the fragment is marginally optically thick (Eqn. \ref{EQN:3DMARG}), and demonstrate that this assumption is valid in the local contemporary interstellar medium. We also derive a simple analytic expression for the luminosity of a spherical fragment which is not necessarily marginally optically thick (Eqns. \ref{EQN:COOL3} \& \ref{EQN:TAU3}), and hence we obtain expressions for the minimum mass which can be used in circumstances other than contemporary local star formation, when it cannot necessarily be assumed a priori that the minimum-mass fragment is marginally optically thick (Eqns. \ref{EQN:3DTHIN} \& \ref{EQN:3DTHICK}). We also rehearse the reasons why hierarchical 3-D fragmentation may not be a useful paradigm for contemporary star formation.

One-shot 2-D fragmentation represents the situation where star formation is triggered by two (or more) supersonically colliding streams, producing a shock-compressed layer (or filament). Therefore (a) the geometry is anisotropic, (b) ram pressure is important, and (c) the fragments are initially flattened. Here the main processes competing against fragmentation -- and thereby fixing the scale of fragmentation -- are the continuing inflow of material into the layer, and the extra heating associated with the resulting accretion shock. The masses of fragments depend on the flux of matter in the colliding flows (Eqn. \ref{EQN:MASSFLUX}), and the initial aspect ratio of a fragment is approximately equal to the Mach Number of the accretion shock (Eqn. \ref{EQN:ASPECT}). Expressions are derived for the time at which the accumulating layer fragments (Eqn. \ref{EQN:tFRAG2}), for the properties of the resulting fragments (Eqns. \ref{EQN:MFRAG2} to \ref{EQN:ZFRAG2}), for the luminosity of a flattened disc-shaped fragment (Eqns. \ref{EQN:COOL2} \& \ref{EQN:TAU2}), and hence for the minimum mass in situations where the fragment is optically thin, marginally optically thick, or strongly optically thick (Eqns. \ref{EQN:2DMARG} to \ref{EQN:2DTHICK}). We also formulate the condition which must be satisfied for minimum-mass fragments to be optically thin or optically thick. Hence we show that minimum-mass fragments in the outer galaxy should be optically thin at their inception, and conversely minimum-mass fragments in the inner galaxy should be optically thick at their inception.

Disc fragmentation represents the situation where the star forming material is in orbit about an existing star and therefore (a) the geometry is again anisotropic, (b) centrifugal acceleration is important. The main process competing with condensation of a proto-fragment is tidal shear, which tends to tear proto-fragments apart, unless they can cool and condense out on a dynamical timescale. We present a derivation of the Toomre criterion which has the advantage both of being very simple (Eqns. \ref{EQN:EVOLD} \& \ref{EQN:ERMOOT}), and of being applicable to non-equilibrium discs. This is important because in real star formation events discs may not have sufficient time to settle into equilibrium before they fragment. In a disc, the masses of fragments depend on the local sound speed, $a$, and the local vorticity, $w$, since condensation of a fragment is resisted by both pressure and rotation. Hence the minimum mass depends on the properties of the central star (mass, $M_\star$, and luminosity, $L_\star$), but with reasonable assumptions this dependence is quite weak.

The fact that, for contemporary local star formation, all three generic scenarios predict a minimum mass in the narrow range $0.001\;{\rm to}\;0.004\,{\rm M}_{_\odot}$, is remarkable. In the different scenarios, the expressions for the Jeans mass are completely different, the geometries are different, the competing processes are different, and the radiation transport regimes are different. In hierarchical 3-D fragmentation, $m_{_{\rm JEANS\,3}}$ depends only on $\rho$ and $a$, and the shortest condensation time is for the largest fragment. In one-shot 2-D fragmentation of a shock-compressed layer, $m_{_{\rm JEANS\,2}}$ depends also on $v$ and $t$, fragments are initially flattened, the critical quantity is the flux of matter flowing into the layer, and the shortest condensation time is for a finite-mass fragment. In a disc, $m_{_{\rm JEANS\,D}}$ depends on $a$ and $\Sigma$, hence on $M_\star$, $L_\star$ and $R$, the critical quantity is $\omega^2/a^5$, and the minimum-mass fragment is optically thick ($\bar{\tau}\sim 100$), unlike in the other two scenarios (where $\bar{\tau}\sim 1$). Yet despite these differences, the minimum mass has a very small range. This implies that in contemporary, local star formation, the details of geometry and background dynamics have little influence on the minimum mass. (It might be tempting, but we believe it would be incorrect, to assert that this result does not need proof.)

In addition to evaluating the minimum mass, we have shown that it is hard for discs to undergo gravitational fragmentation at small radii, basically because the timescale on which a proto-fragment is sheared apart is shorter than the timescale on which it can cool and condense out. Consequently there should be a dearth of brown dwarfs in orbit around Sun-like stars, and this provides a possible explanation for the observationally inferred Brown Dwarf Desert. Conversely, brown dwarfs can condense out at large radii in massive extended discs, because here the timescale for a proto-fragment to be sheared apart is likely to exceed the cooling timescale. We therefore suggest that disc fragmentation is the most likely formation mechanism for those brown dwarfs seen in wide orbits about Sun-like stars. We also point out that fragmentation of the outer parts of massive extended discs should occur very quickly, essentially on a dynamical timescale ($\sim 10^4\,{\rm years}$), and therefore the paucity of massive extended discs need not mean that they do not form, but simply that they quickly self-destruct by condensing into brown dwarfs and low-mass H-burning stars.

\subsection{Secondary Fragmentation during H$_{_2}$ dissociation}

A prestellar condensation formed by Primary Fragmentation can fragment further during the Second Collapse, when dissociation of H$_{_2}$ acts as a sink for the internal energy being delivered by compression. If at this stage the protostellar condensation is flattened by rotation, the Secondary Collapse can lead to Secondary Fragmentation and the formation of a close low-mass binary system. From simple thermodynamic arguments, H$_{_2}$ dissociation occurs at $T_{_{\rm DISS}} \sim 2000\,{\rm K}$ and $\rho_{_{\rm DISS}} \sim 10^{-7}\,{\rm g}\,{\rm cm}^{-3}$. Therefore the resulting binary separations should approximate to
\begin{eqnarray}
a&\sim&5\,{\rm AU}\,\left[\frac{M_{_{\rm SYSTEM}}}{0.1\,{\rm M}_{\odot}}\right]\,.
\end{eqnarray}
If this is compared with the plot of Close et al. (2003; their Fig. 15), it is seen to provide a remarkably good fit to the observed separations of close low-mass systems, suggesting strongly that these systems have been formed by Secondary Fragmentation.

Since Secondary Fragmentation requires that the primary fragment be rotating, and since the only fragmentation scenario we have considered which explicitly takes account of the spin of a proto-fragment is disc fragmentation, it is tempting to consider the outer parts of discs as the location for forming close low-mass binary systems. This suggestion derives support from the observation that brown dwarfs in wide orbits about Sun-like stars appear more likely to have brown-dwarf companions than brown dwarfs in the field (Burgasser et al. 2005). The simplest explanation for this would seem to be that low-mass prestellar condensations form by Primary Fragmentation in the outer parts of discs (as described in Section \ref{SEC:DISC}) and then undergo Secondary Fragmentation to form close low-mass binaries, in situ.

Furthermore we show that close low-mass binaries formed in this way can sometimes survive being impulsively separated from the Sun-like star, and therefore brown-dwarf binaries in the field could also have formed by disc fragmentation.

\appendix

\section{Radiation transport in a one-zone fragment}

\subsection{A spherical fragment}

To justify both Eqns. (\ref{EQN:COOL3}) and (\ref{EQN:TAU3}), we consider a uniform-density spherical fragment of mass $m$ and radius $r$, and treat separately the optically thin and optically thick limits.

In the optically thin limit, the term $\bar{\tau}_{_{\rm R}}(T)$ can be neglected in comparison with $\bar{\tau}_{_{\rm P}}^{\;-1}(T)$, and so Eqn. (\ref{EQN:COOL3}) reduces to
\begin{eqnarray}
{\cal L}_{_{\rm THIN}} & =  & 4\,m\,\bar{\kappa}_{_{\rm P}}(T)\,\sigma_{_{\rm SB}}\,T^4 \,.
\end{eqnarray}
This is an exact expression, in the sense that it defines the Planck-mean opacity, $\bar{\kappa}_{_{\rm P}}(T)$.

In the optically thick limit, the term $\bar{\tau}_{_{\rm P}}^{\;-1}(T)$ can be neglected in comparison with $\bar{\tau}_{_{\rm R}}(T)$, and so Eqn. (\ref{EQN:COOL3}) now reduces to
\begin{eqnarray} \label{EQN:THICK1}
{\cal L}_{_{\rm THICK}} & \simeq & \frac{16\,\pi\,r_{_3}^2\,\sigma_{_{\rm SB}}\,T^4}{3\,\bar{\tau}_{_{\rm R}}(T)} \,.
\end{eqnarray}
Since $\bar{\tau}_{_{\rm R}}(T)$ is the Rosseland-mean optical-depth between the centre of the spherical fragment and its surface, we can define the diffusion time,
\begin{eqnarray}
t_{_{\rm DIFF}} & \simeq & \frac{r_{_3}\,\bar{\tau}_{_{\rm R}}(T)}{c}\,,
\end{eqnarray}
which is just the time it takes radiation to random-walk from the interior of the spherical fragment to its exterior. Eliminating $\bar{\tau}_{_{\rm R}}(T)$ in favour of $t_{_{\rm DIFF}}$, and putting $\sigma_{_{\rm SB}} = a_{_{\rm SB}} c / 4$ (where $a_{_{\rm SB}}$ is the radiation energy-density constant), Eqn. (\ref{EQN:THICK1}) becomes
\begin{eqnarray} \label{EQN:THICK2}
{\cal L}_{_{\rm THICK}} & \simeq & \frac{4\,\pi\,r_{_3}^3\,a_{_{\rm SB}}\,T^4\,/\,3}{t_{_{\rm {\rm DIFF}}}} \,.
\end{eqnarray}
The righthand side of Eqn. (\ref{EQN:THICK2}) is the total radiant energy inside the spherical fragment divided by the time this radiation takes to escape. In the optically thick limit this is a reasonable approximation to the luminosity.

Finally we note that, when the spherical fragment is neither optically thin, nor optically thick, and $\bar{\tau}_{_{\rm R}}(T) \simeq \bar{\tau}_{_{\rm P}}(T) \simeq 1$, the luminosity of the fragment is given by
\begin{eqnarray}
{\cal L}_{_{\tau \sim 1}} & \simeq & 4\,\pi\,r_{_3}^2\,\sigma_{_{\rm SB}}\,T^4 \;\times\; \frac{2}{3} \,,
\end{eqnarray}
so it cools almost as well as a blackbody.

\subsection{A disc-like fragment}

To justify both Eqns. (\ref{EQN:COOL2}) and (\ref{EQN:TAU2}) --- and Eqns. (\ref{EQN:LDISC}) and (\ref{EQN:tauDISC}), with $r_{_2}\!\rightarrow\!r_{_{\rm DISC\,D}}$ --- we consider a uniform-density disc-like fragment of mass $m_{_2}$, radius $r_{_2}$, and thickness $2z_{_2}$ (where the assumption is that $z \ll r$, i.e. the disc is geometrically thin). We again treat separately the optically thin and optically thick limits.

In the optically thin limit, the term $\bar{\tau}_{_{\rm R}}(T)$ can be neglected in comparison with $\bar{\tau}_{_{\rm P}}^{\;-1}(T)$, and so Eqn. (\ref{EQN:COOL2}) reduces to the exact form:
\begin{eqnarray}
{\cal L}_{_{\rm THIN}} & = & 4\,m_{_2}\,\bar{\kappa}_{_{\rm P}}(T)\,\sigma_{_{\rm SB}}\,T^4 \,.
\end{eqnarray}

In the optically thick limit, the term $\bar{\tau}_{_{\rm P}}^{\;-1}(T)$ can be neglected in comparison with $\bar{\tau}_{_{\rm R}}(T)$, and so Eqn. (\ref{EQN:COOL2}) now reduces to
\begin{eqnarray} \label{EQN:THICK3}
{\cal L}_{_{\rm THICK}} & \simeq & \frac{4\,\pi\,r_{_2}^2\,\sigma_{_{\rm SB}}\,T^4}{\bar{\tau}_{_{\rm R}}(T)} \,.
\end{eqnarray}
The time for radiation to diffuse from the interior of the disc-like fragment to its flat surfaces at $\pm z$ is
\begin{eqnarray}
t_{_{\rm DIFF}} & \simeq & \frac{2\,z_{_2}\,\bar{\tau}_{_{\rm R}}(T)}{c}\,,
\end{eqnarray}
where the factor 2 arises because the radiation must diffuse in the $z$-direction to escape. It follows that Eqn. (\ref{EQN:THICK3}) can be rewritten as
\begin{eqnarray} \label{EQN:THICK4}
{\cal L}_{_{\rm THICK}} & \simeq & \frac{2\,\pi\,r_{_2}^2\,z_{_2}\,a_{_{\rm SB}}\,T^4}{t_{_{\rm {\rm DIFF}}}} \,.
\end{eqnarray}
The righthand side of Eqn. (\ref{EQN:THICK4}) is the total radiant energy inside the disc-like fragment, divided by the time this radiation takes to escape, which in the optically thick limit is a reasonable approximation to the luminosity.

When the disc-like fragment is neither optically thin, nor optically thick, and $\bar{\tau}_{_{\rm R}}(T) \simeq \bar{\tau}_{_{\rm P}}(T) \simeq 1$, its luminosity is given by
\begin{eqnarray}
{\cal L}_{_{\tau \sim 1}} & \simeq & 2\,\pi\,r_{_2}^2\,\sigma_{_{\rm SB}}\,T^4 \;\times\; 1 \,,
\end{eqnarray}
so it cools exactly like a black body from its flat surfaces.

\begin{acknowledgements}
We gratefully acknowledge the support of PPARC Grant PPA/G/O/2002/00497.
\end{acknowledgements}

\end{document}